\documentclass[conference]{IEEEtran}
\newif\iffinal
\usepackage{algorithm}

\def\eqref#1{Equation~\ref{#1}}
\usepackage{alltt}          % add typewriter type

\usepackage[normalem]{ulem} % for strikeout (\sout) and squiggly lines (\uwave)
\usepackage{nicefrac}       % compact symbols for 1/2, etc.

\usepackage{hyperref}
\usepackage[geometry]{ifsym}
\usepackage{array}
\usepackage{comment}
\usepackage[us]{datetime}
\usepackage{environ}
\usepackage{float}
\usepackage{framed}
\usepackage{graphicx}
\usepackage{siunitx}
\usepackage{microtype}

\usepackage{url}
\usepackage{verbatim}
\usepackage{wrapfig}

\usepackage{pifont} % for ding numbers

\usepackage{framed}
\usepackage[at]{easylist} % for concise and nested lists
\usepackage{paralist}
\usepackage{enumitem}

\usepackage{tabularx}
\usepackage{rotating}           %mainly for rotating tables
\usepackage{booktabs}
\usepackage{longtable}

\usepackage{pdfpages}

\ifthenelse{\isundefined{\email}}{
    \newcommand{\email}[1] {\texttt{\small\href{mailto:#1}{#1}}}
}{
}

\newcommand{\smartincludetex}[1]{
    \IfFileExists{#1.tex}{
        \input{#1}
    }{
        \iffinal
        \else
        \critfix{No file found for '#1'!}
        \fi
        \begin{center}
            \textbf{\color{red}\Large No file found for '#1.tex'!} \\
        \end{center}
        \typeout{smartinclude: WARNING: File not found: #1.tex}
    }
}

\definecolor{limeyellow}{RGB}{228,255,100}
\usepackage[colorinlistoftodos,prependcaption,textsize=footnotesize,color=limeyellow,linecolor=magenta]{todonotes}

\iffinal
    
    \newcommand{\ar}[1]{} % Action Required
    \newcommand{\guidance}[1]{}
    \newcommand{\improvement}[1]{}
    \newcommand{\info}[1]{}
    \newcommand{\notes}[2]{}
    \newcommand{\poc}[1]{}
    \newcommand{\secparam}[1]{}
    \newcommand{\unsure}[1]{}
    
    \newcommand{\jv}[1]{}
    \newcommand{\nr}[1]{}
    \newcommand{\fyl}[1]{}
    \newcommand{\ay}[1]{}

    \NewEnviron{rfpinfo}
    {\begin{tcolorbox}[colback=blue!5!white,colframe=blue!75!black]\noindent\sffamily\footnotesize
    \end{tcolorbox}}

\else
    \newcommand{\ar}[1]{\todo[inline,linecolor=blue,backgroundcolor=blue!25,bordercolor=blue,textcolor=black]{#1}}
    \newcommand{\guidance}[1]{\todo[inline,backgroundcolor=yellow,textcolor=black,size=footnotesize]{#1}}
    \newcommand{\improvement}[1]{\todo[inline,linecolor=Plum,backgroundcolor=Plum!25,bordercolor=Plum]{#1}}
    \newcommand{\info}[1]{\todo[inline,linecolor=PineGreen,backgroundcolor=PineGreen!25,bordercolor=PineGreen]{#1}}
    \newcommand{\notes}[2]{\textcolor{magenta}{\sf {\small~\$\$ [#1] #2}}}
    \newcommand{\poc}[1]{\textcolor{teal}  {\sf {\footnotesize~ @#1}}}
    \newcommand{\secparam}[1]{\todo[inline,linecolor=gray,bordercolor=gray,backgroundcolor=lightgray,textcolor=black]{#1}}
    \newcommand{\unsure}[1]{\todo[inline,linecolor=red,backgroundcolor=red!25,bordercolor=red]{#1}}

	\newcommand{\jv}[1]{\ar{[JV] #1}}
    \newcommand{\nr}[1]{\ar{[NR] #1}}
    \newcommand{\fyl}[1]{\ar{[FYL] #1}}
	\newcommand{\ay}[1]{\ar{[AY] #1}}
	
	\NewEnviron{rfpinfo}
	{\begin{tcolorbox}[colback=blue!5!white,colframe=blue!75!black]\noindent\sffamily\footnotesize
			\BODY
	\end{tcolorbox}}

\fi

\newcommand{\critfix}[1]{\todo[inline]{~CRIT: #1}}
\newcommand{\PUNT}[1]{}                   

\usepackage{svg}
\usepackage{fourier}
\usepackage{tikz}
\usepackage{circuitikz}
\usetikzlibrary{arrows.meta}
\usetikzlibrary{tikzmark}
\usetikzlibrary{calc}
\usetikzlibrary{positioning}
\usepackage{tcolorbox}
\tcbuselibrary{skins}

\IEEEoverridecommandlockouts
\usepackage{cite}
\usepackage{amsmath,amssymb,amsfonts}
\usepackage{algorithmic}
\usepackage{graphicx}
\usepackage{textcomp}
\usepackage{xcolor}
\usepackage{array}
\usepackage{subcaption}
\usepackage{etoolbox}
\usepackage{cleveref}

\makeatletter
\newcommand{\linebreakand}{%
  \end{@IEEEauthorhalign}
  \hfill\mbox{}\par
  \mbox{}\hfill\begin{@IEEEauthorhalign}
}
\makeatother

\def\BibTeX{{\rm B\kern-.05em{\sc i\kern-.025em b}\kern-.08em
    T\kern-.1667em\lower.7ex\hbox{E}\kern-.125emX}}

\usepackage{anyfontsize}

\usepackage{atbegshi,multido}
\makeatletter
\newcommand{\removepage}[1]{%
  \def\@drop@this@page{\AtBeginShipoutDiscard}%
  \multido{\iA=1+1}{#1}{%
    \expandafter\gdef\expandafter\@drop@this@page\expandafter%
      {\expandafter\AtBeginShipoutNext\expandafter{\@drop@this@page}}%
    }%
  \@drop@this@page%
}

\begin{document}

% Hook into references.bib and execute the @IEEEtranBSTCTL
% block, enforcing the DATE conference paper rule:
%     Unless there are six authors or more give all authors' names
\bstctlcite{DATEetal:BSTcontrol}

\removepage{8}

\title{
% Tweak first value until max font size for 2 lines
\fontsize{0.0327\textheight}{0.05\textheight}\selectfont
Mapping Spiking Neural Networks to Heterogeneous Crossbar Architectures using Integer Linear Programming}

% Blind authors? True => Replace with anonymizing informaiton
\newtoggle{anon_authors}
\togglefalse{anon_authors}

% Abstract only? True => Only title, authors, and abstract will be included.
\newtoggle{abstract_only}
\togglefalse{abstract_only}

% Tex hacking
\iftoggle{anon_authors}{
\newcounter{authorCount}
\let\oldIEEEauthorblockN\IEEEauthorblockN
\renewcommand{\IEEEauthorblockN}[1]{\oldIEEEauthorblockN{Author \arabic{authorCount}}}
\let\oldIEEEauthorblockA\IEEEauthorblockA
\renewcommand{\IEEEauthorblockA}[1]{\oldIEEEauthorblockA{\textit{Affiliation \arabic{authorCount}}\\email@domain.org}\stepcounter{authorCount}}
\setcounter{authorCount}{-9} % Correct for tex unfurling nonsense
}

% Now for the author block
% If anon_authors is set, this will contain procedurally generated nonsense

\author{\IEEEauthorblockN{Devin Pohl}
\IEEEauthorblockA{\textit{Georgia Institute of Technology}\\
dpohl@gatech.edu}
\and
\IEEEauthorblockN{Aaron Young}
\IEEEauthorblockA{\textit{Oak Ridge National Lab}\\
youngar@ornl.gov}
\and
\IEEEauthorblockN{Kazi Asifuzzaman}
\IEEEauthorblockA{\textit{Oak Ridge National Lab}\\
asifuzzamank@ornl.gov}
\linebreakand
\IEEEauthorblockN{Narasinga Rao Miniskar}
\IEEEauthorblockA{\textit{Oak Ridge National Lab}\\
miniskarnr@ornl.gov}
\and
\IEEEauthorblockN{Jeffrey S. Vetter}
\IEEEauthorblockA{\textit{Oak Ridge National Lab}\\
vetter@ornl.gov}
}
% First 2-3 entences frame problem. and what we're tyring to solve
% Work of the paper is about half way through the abstract
%
% 

% When using two author lines, ieeetran ends up putting too much extra space after
% the author block. The below removes that space to make the document look more
% consistant (and give us a tiny bit more room to write).
\IEEEaftertitletext{\vspace{-0.95\baselineskip}}

% Lookup for ranges
\newcommand{\rangeof}[1]{%
  \ifthenelse{\equal{#1}{area_hom}}{16.7--27.6\%}{%
  \ifthenelse{\equal{#1}{breakeven_hom}}{2.5--13.2x}{%
  \ifthenelse{\equal{#1}{area_het_over_hom}}{66.9--72.7\%}{%
  \ifthenelse{\equal{#1}{breakeven_het}}{0.15--3.73x}{%
  \ifthenelse{\equal{#1}{snu}}{11.9--26.4\%}{%
  \ifthenelse{\equal{#1}{snu_hom}}{9.2--26.9\%}{%
  \ifthenelse{\equal{#1}{pgo}}{0.5--14.8\%}{%
  \ifthenelse{\equal{#1}{pgo_magnitude_speedup}}{1--3}{
  Key not found}}}}}}}}}%

\maketitle

\begin{abstract}
Advances in novel hardware devices and architectures allow Spiking Neural Network (SNN) evaluation using ultra-low power, mixed-signal, memristor crossbar arrays.
As individual network sizes quickly scale beyond the dimensional capabilities of single crossbars, networks must be mapped onto multiple crossbars.
Crossbar sizes within modern Memristor Crossbar Architectures (MCAs) are determined predominately not by device technology but by network topology; more, smaller crossbars consume less area thanks to the high structural sparsity found in larger, brain-inspired SNNs.
Motivated by \textit{continuing} increases in SNN sparsity due to improvements in training methods, we propose utilizing heterogeneous crossbar sizes to further reduce area consumption.
This approach was previously unachievable as prior compiler studies only explored solutions targeting homogeneous MCAs.
Our work improves on the state-of-the-art by providing Integer Linear Programming (ILP) formulations supporting arbitrarily heterogeneous architectures.
By modeling axonal interactions between neurons, our methods produce better mappings while removing inhibitive a priori knowledge requirements.
We first show a \rangeof{area_hom} reduction in area consumption for square-crossbar homogeneous architectures.
Then, we demonstrate \rangeof{area_het_over_hom} \textit{further} reduction when using a reasonable configuration of heterogeneous crossbar dimensions.
Next, we present a new optimization formulation capable of minimizing the number of inter-crossbar routes.
When applied to solutions already near-optimal in area, an \rangeof{snu} routing reduction is observed without impacting area consumption.
Finally, we present a profile-guided optimization capable of minimizing the number of runtime spikes between crossbars.
Compared to the best-area-then-route optimized solutions, we observe a further \rangeof{pgo} inter-crossbar spike reduction while requiring \rangeof{pgo_magnitude_speedup} orders of magnitude less solver time.
\end{abstract}

%\begin{IEEEkeywords}
%component, formatting, style, styling, insert
%\end{IEEEkeywords}

\iftoggle{abstract_only}{}{
% Total 6 pages

\section{Introduction}

\looseness-1
%The forefront of hardware-accelerated machine learning is currently experiencing a pivotal moment.
%The near-universal application of neural networks has outgrown traditional models (DNNs, CNNs, etc.) under expanding application spaces and problem complexity, with Spiking Neural Networks (SNNs) gaining popularity thanks to comparable accuracy despite significantly lower neuron counts \cite{IrizarryValle2015AnAN,SnnAreEfficient,SnnAreAccurate1,SnnAreAccurate2,SnnAreAccurate3}.
%Simultaneously, co-design with resistive-RAM technologies has produced material, device, and architecture-level improvements to Memristor Crossbar Architectures (MCAs)
%Together, these advances unlock remarkably power- and area-efficient neuromorphic computing \cite{ReRamOrigins,future:2018:zidan,9365958,9062958,9366045,NeuroCIM}.
%For the first time, co-design between MCA-based accelerators and accelerator-tailored SNNs is approaching a full-stack solution for accessible, ultra-low-power inference of general-purpose, highly-accurate machine learning models.
The forefront of hardware-accelerated machine learning is currently experiencing a pivotal moment.
The near-universal application of neural networks has outgrown traditional models (DNNs, CNNs, etc.) under expanding application spaces and problem complexity, with Spiking Neural Networks (SNNs) gaining popularity thanks to comparable accuracy despite significantly lower neuron counts \cite{IrizarryValle2015AnAN,SnnAreEfficient,SnnAreAccurate1,SnnAreAccurate2,SnnAreAccurate3}.
Simultaneously, co-design with resistive-RAM technologies has contributed to material, device, and architecture-level improvements to Memristor Crossbar Architectures (MCAs).
These developments together enable remarkably power- and area-efficient neuromorphic computing \cite{ReRamOrigins,future:2018:zidan,9365958,9062958,9366045,NeuroCIM}.
For the first time, co-design between MCAs and SNNs is nearing a full-stack solution for accessible, ultra-low-power inference of general-purpose, highly-accurate machine learning models.

%However, due to improvements in training methods, dominant network structures of SNNs have shifted in recent years.
%Between advances in initial network generation \cite{shang2020conversion,SNNMoreSparse,direct_training}, training \cite{markram2011history,nunes2022spiking}, and pruning methods \cite{SNNMoreSparse,10387103,GONG2024127059,han2023adaptivesparsestructuredevelopment}, structural sparsity has increased drastically.
%Where previously, architectures only needed to support a large number of small crossbars to support ever-growing networks, today's architectures require ever more ingenious tricks to fully take advantage of increasing sparsity \cite{9365958,9062958,9366045,NeuroCIM}.
However, dominant network structures of SNNs have shifted in recent years.
Between advances in initial network generation \cite{shang2020conversion,SNNMoreSparse,direct_training}, training \cite{markram2011history,nunes2022spiking}, and pruning methods \cite{SNNMoreSparse,10387103,GONG2024127059,han2023adaptivesparsestructuredevelopment}, networks have evolved drastically increased structural sparsity.
Where previously, architectures only needed many small crossbars to support ever-growing networks, today's architectures require ever more ingenious tricks to fully take advantage of increasing sparsity \cite{9365958,9062958,9366045,NeuroCIM}.

%Yet, these architectural improvements outpace compiler technology.
%The task of partitioning SNNs and mapping neuron clusters to different crossbars is very much an open problem.
%Many recent approaches scale with increasing network size \cite{RecentApproach4,RecentApproach3,RecentApproach1,RecentApproach2},
% Note: The order of the above reference list is very imporant!
% These references end up right at the page break of the references
% page, and we need careful control to keep references on one page.
%but none support the level of target architecture heterogeneity required to leverage structural sparsity for improved area and power metrics.
%To the best of our knowledge, SpikeHard\cite{spikehard} is the first study with potential to jump the gap, utilizing Integer Linear Programming (ILP) for improved area optimization.

Yet, these architectural improvements outpace compiler technology.
Partitioning SNNs and mapping neuron clusters to various crossbars remains an open problem, with many recent approaches scaling to increasing network sizes \cite{RecentApproach4,RecentApproach3,RecentApproach1,RecentApproach2}.
% Note: The order of the above reference list is very imporant!
% These references end up right at the page break of the references
% page, and we need careful control to keep references on one page.
However, these approximate solutions fail to support the level of target architecture heterogeneity required for leveraging structural sparsity to improve area and power metrics.
To the best of our knowledge, SpikeHard\cite{spikehard} is the first study with potential to jump the gap, utilizing Integer Linear Programming (ILP) for improved area optimization.

However, two main limitations are present with this work: (1) it \textit{requires} an initial solution and (2) it produces solutions demonstrably sub-optimal in area consumption.
Furthermore, the base ILP constraints presented in SpikeHard prohibit the optimization of more useful heuristics.

This work improves on the state-of-the-art in SNN to MCA mapping through the following contributions:
\begin{itemize}
\item Providing ILP formulations that remove the need for an a priori known-valid solution while allowing for production of truly area-optimal solutions.
\item Providing ILP formulations to minimize the number of required network routes between crossbars.
\item Providing ILP formulations leveraging prior inference spike profiles for minimizing runtime network packets.
\end{itemize}

\noindent
The rest of this text is organized as follows: Section \ref{sec:background} discusses background material, section \ref{sec:relatedwork} explains opportunities for improvement over the state-of-the-art, section \ref{sec:approach} shows the contributed ILP formulations and their meaning, and section \ref{sec:results} experimentally shows the effectiveness and tradeoff characteristics of these techniques.
\section{background}\label{sec:background}
{% Layout: This needs to fit in 4 lines, but I've already cut out as much as possible without it sounding weird. So let's cheat a bit:
\looseness=-1 \spaceskip= 2pt plus 1pt minus 1.5pt  \spaceskip= 3pt plus 2pt minus 2pt
This section summarizes key concepts and trends including spiking neural networks and training methods, memristor crossbar architectures and scaling efforts, use of integer linear programming in mapping, and profile-guided optimization.}

\subsection{Spiking Neural Networks}
Neuromorphic computing involves utilizing bio-plausible artificial neural networks for computation.
While this can include emulating neural components beyond just neurons and axons \cite{IrizarryValle2015AnAN}, the approach most relevant to applied machine learning is the Spiking Neural Network (SNN) with simple integrate-and-fire neurons.
%In SNNs, neurons accumulate charge and, upon reaching a threshold, fire discrete "spikes" of information along axons which may delay the signals---rather than constantly executing mathematical functions as in DNNs, GNNs, etc.
Instead of executing every neuron per input window like in DNNs, GNNs, etc, neurons in SNNs accumulate charge and, upon reaching a threshold, fire discrete "spikes" of information along axons which may delay the signals.
With appropriate hardware, SNNs achieve much higher energy efficiency without significant loss in accuracy \cite{SnnAreEfficient,SnnAreAccurate1,SnnAreAccurate2,SnnAreAccurate3}.

Training SNNs is still an open problem with fervent development \cite{schuman2022opportunities}.
While one class of training methods focuses on converting other types of networks (mainly CNNs, but also DNNs) into SNNs through various processes \cite{9597482,10387103,10.3389/fnins.2020.00439,10.3389/fnins.2017.00682,cao2015spiking}, literature suggests that training SNNs "from scratch" produces networks with higher amounts of certain properties (e.g., gradient sparsity) \cite{direct_training,SNNMoreSparse}.
Continuing research on pruning, compression, and re-structuring SNNs show increasing ability to take advantage of these properties to produce more structurally sparse networks \cite{SNNMoreSparse,10387103,GONG2024127059,han2023adaptivesparsestructuredevelopment}.

\subsection{Memristor Crossbar Architectures}
Resistive-RAM (ReRAM) offers non-volatile random access memory while also supporting analog compute.
While ReRAM research is not new \cite{ReRamOrigins}, recent advances in memristor technology and architecture \cite{ReRamRecent} rapidly approach production-ready systems for accelerating general-purpose SNNs.
Analog ReRAM crossbar-based accelerators enable energy-efficient matrix multiplication but face non-idealities that limit crossbar dimensions.

Scaling architectures to network size (neuron count) is achieved by minimizing connection overhead via methods such as mixed-signal accumulation and hierarchical networking \cite{9365958,9062958,9366045}.
Scaling to density (edge count) requires more intricate tricks, such as re-purposing crossbar bit-lines for metadata \cite{NeuroCIM}.
Yet, the corresponding compiler technology to fully exploit such architectures remains notably absent.

\subsection{Integer Linear Programming}
The architectural advantages of high crossbar counts come with significant compiler challenges; after all, complex architectures are only useful if compilers can effectively leverage their complexity. For NP-hard problems like SNN-to-MCA mapping, traditional wisdom favors approximate, polynomial-time algorithms \cite{RecentApproach1,RecentApproach2,RecentApproach3,RecentApproach4}. While these methods provide \textit{adequate} solution quality, they fall short in supporting emerging heterogeneous crossbar architectures.

An alternative approach is Integer Linear Programming (ILP), a type of constraint programming requiring only mathematical descriptions of valid solutions.
Off-the-shelf solvers then find satisfying and/or optimal solutions within generally tolerable time limits, despite NP-hard problem complexity.
While recent work leverages ILP to map SNNs to MCAs with impressive reductions in area consumption \cite{spikehard}, we will show further opportunities for reducing area consumption and inter-crossbar communication.

\subsection{Profile Guided Optimization}
The methods in this paper, in part, leverage Profile Guided Optimization (PGO) to improve average-case performance.
PGO involves sampling execution to identify frequently activated components and optimizing them more aggressively.

In the context of spiking networks, certain synapses consistently experience more spikes across varying input patterns \cite{daniel_j__amit__2003,pastorelli2019scaling,bartram2023parallel}.
By clustering these "hot" synapses within single crossbars, expensive inter-crossbar communication is only required for infrequently utilized routes.
\section{Related Work}\label{sec:relatedwork}

Previous approaches to this mapping problem include block clustering \cite{RecentApproach4}, spectral clustering \cite{RecentApproach2}, exclusive sum-of-product mapping \cite{RecentApproach3}, and sum-of-cut-cost partitioning \cite{RecentApproach1}.
While effective for scaling to large network sizes, these methods produce sub-optimal mappings \cite{spikehard} and only support homogeneous MCAs.
Modifying such algorithms for heterogeneous MCA support is yet to be attempted.

To our knowledge, only one work has employed ILP as an alternative method: SpikeHard \cite{spikehard}.
This approach groups neurons into \textit{Minimally Connected Components} (MCCs) and uses their aggregate dimension requirements for bin-packing.
The greatest limitation is that it requires an initial solution to form MCCs---forming single-neuron MCCs is disastrous for optimization and cannot be worked around via multiple SpikeHard applications, shown empirically in Section \ref{sec:area_compare}.

The second limitation is that area-optimized results are sub-optimal.
This approach does not model axon-sharing, where a single word-line supplies input to multiple bit-lines in a crossbar.
Consequently, placing two MCCs in the same crossbar may incorrectly require additional input lines. 
This effect is shown in Fig. \ref{fig:mcc} and is addressed in our approach.

The final limitation with SpikeHard is its lack of support for more complex optimizations.
With input axons counted incorrectly, neither inter-crossbar connections nor network weights can be modeled with reasonable accuracy.

\begin{figure}[b]
\vspace{-1.7\baselineskip}
    \centering
    \begin{tikzpicture}[scale=0.52, every node/.style={scale=0.92}]
    \draw[thick] (0,0) rectangle (4.5, 4.5);

    % Spacing for parallel lines
    \pgfmathsetmacro{\offset}{10}

    \draw (-0.5, 4) -- (3.5, 4);
    \draw (-0.5, 3) -- (3.5, 3);
    \draw (-0.5, 2) -- (3.5, 2);
    
    \draw (1, 4.5) -- (1, 1);
    \draw (1, 0.25) -- (1, -0.25);
    \draw (2, 4.5) -- (2, 1);
    \draw (2, 0.25) -- (2, -0.25);
    \draw (3, 4.5) -- (3, 1);
    \draw (3, 0.25) -- (3, -0.25);

    \foreach \j [evaluate=\j as \y using \j*2] in {1,2,3} {
      \foreach \i [evaluate=\i as \x using \i*2] in {1,2,3} {
        \draw (0, 0) [scale=0.5] (\x, 8.5-\y) to [memristor, transform shape] (\x-1.5, 10-\y);
      }
    }

    \draw (2.8,0.65) node[above,rotate=90] {\huge...\normalsize};
    \draw (4,1.2) node[above] {\huge...\normalsize};

    \draw (-0.75, 4) node[] {\textbf{1}};
    \draw[-,blue,line width=1mm] (-1.1, 4) ++(0, 0.3) -- ++(0, -0.6);
    \draw (-0.75, 3) node[] {\textbf{2}};
    \draw[-,green,line width=1mm] (-1.1, 3) ++(0, 0.3) -- ++(0, -0.6);
    \draw (-0.75, 2) node[] {\textbf{2}};
    \draw[-,purple,line width=1mm] (-1.1, 2) ++(0, 0.3) -- ++(0, -0.6);
    
    \draw (1, -0.25) node[below] {\textbf{3}};
    \draw (2, -0.25) node[below] {\textbf{4}};
    \draw (3, -0.25) node[below] {\textbf{5}};

    %\draw[red, line width=0.75mm]   (-0.75,3.5) ellipse (0.5cm and 0.875cm);
    %\draw[cyan, line width=0.75mm] (-0.75,2) ellipse (0.5cm and 0.475cm);
    
    \draw[red, line width=0.75mm]   (1.5,-0.675) ellipse (0.875cm and 0.5cm);
    \draw[cyan, line width=0.75mm] (3,-0.675) ellipse (0.475cm and 0.5cm);

    %%%
    \draw[-{Triangle[width=18pt,length=8pt]}, line width=10pt](-4.25,1.75) -- ++(2, 0);
    %%%

    \draw (-7, 1.75) coordinate (left);

    \node[draw, circle] (n4) at (left) {\textbf{4}};
    \node[draw, circle] (n5) at ($(n4)+(0, -2.0)$) {\textbf{5}};
    \node[draw, circle] (n3) at ($(n4)+(0,  2.0)$) {\textbf{3}};
    
    \draw[red, line width=0.75mm]  ($(n3)!0.5!(n4)$) ellipse (1.0cm and 1.75cm);
    \draw[cyan, line width=0.75mm] (n5) ellipse (0.75cm and 0.75cm);

    \node[red,right] at ($(n3)+(0.75, 0.525)$) {MCC 0};
    \node[cyan,right] at ($(n5)+(0.75, -0.525)$) {MCC 1};

    \node[draw, circle] (n1) at ($(n3)+(-3,0)$) {\textbf{1}};
    \node[draw, circle] (n2) at ($(n4)!0.5!(n5)+(-3,0)$) {\textbf{2}};

    \draw[-Triangle,line width=0.35mm] (n1) -- (n3)  coordinate[pos=0.4] (midpoint) coordinate[at end] (end) coordinate[at start] (start);
    \path let \p1=(start),\p2=(end), \n1={veclen(\y2-\y1, -\x2+\x1)},\n2={(\y2-\y1)/\n1*\offset},\n3={(-\x2+\x1)/\n1*\offset} in coordinate (normal) at (\n2, \n3);
    \draw[-,blue,line width=1mm] ($(midpoint)+(normal)$) -- ($(midpoint)-(normal)$);

    \draw[-Triangle,line width=0.35mm] (n2) -- (n4)  coordinate[pos=0.4] (midpoint) coordinate[at end] (end) coordinate[at start] (start);
    \path let \p1=(start),\p2=(end), \n1={veclen(\y2-\y1, -\x2+\x1)},\n2={(\y2-\y1)/\n1*\offset},\n3={(-\x2+\x1)/\n1*\offset} in coordinate (normal) at (\n2, \n3);
    \draw[-,green,line width=1mm] ($(midpoint)+(normal)$) -- ($(midpoint)-(normal)$);
    
    \draw[-Triangle,line width=0.35mm] (n2) -- (n5)  coordinate[pos=0.4] (midpoint) coordinate[at end] (end) coordinate[at start] (start);
    \path let \p1=(start),\p2=(end), \n1={veclen(\y2-\y1, -\x2+\x1)},\n2={(\y2-\y1)/\n1*\offset},\n3={(-\x2+\x1)/\n1*\offset} in coordinate (normal) at (\n2, \n3);
    \draw[-,purple,line width=1mm] ($(midpoint)+(normal)$) -- ($(midpoint)-(normal)$);

    \node[above] (la) at (2.25, 4.55) {\underline{Crossbar}};
    \node[above] at (-7.5, 4.55) {\underline{Network}};

    \node[left,scale=1.5] (xx) at (-2, 3.5) {\warning};
    \draw[-Triangle, dashed] (xx) -- (-1.4, 3.1);
    \draw[-Triangle, dashed] (xx) -- (-1.4, 2.3);

\end{tikzpicture}

    \vspace*{-0.8em} % Phantom out the output ellipses

    \caption{MCC Packing Causing Multiple Counting of Axons
    \vspace{-0.41em}} % Squeeeeeeeeze
    \label{fig:mcc}
\end{figure}
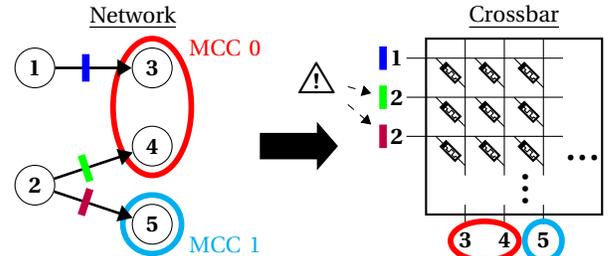

%%%%%%%%%%%%%%

\section{Approach}\label{sec:approach}
This work moves away from traditional approximate solutions which sacrifice either optimality \cite{RecentApproach1} or accuracy \cite{AffectsAccuracy}. 
We first modify ILP constraints from \cite{spikehard} to support axon sharing, yielding more optimal area consumption and enabling advanced optimizations.
We then apply an optimization for minimizing the count of inter-crossbar networking routes.
Lastly, we introduce a profile-guided optimization to reduce runtime packets across chip routers.

\subsection{Formulation of Constraints}
The root cause of SpikeHard's shortcomings---incorrectly counting shared axons as in Fig. \ref{fig:mcc}---is overcome with additional neuron placement variables.
The high-level method of constructing the set of axons mapped as input to a given crossbar $j$ is as follows:
\begin{equation}
\text{Inputs}_j = \bigcup_{i \in \text{Outputs}_j} \text{InputEdgesOfNode}(i) 
\end{equation}

\noindent
This can be re-written imperatively in boolean logic by introducing $x_{ij}$ for neuron placement and $s_{kj}$ as axon placement.
With graph edge definitions stored in $m_{ik}$:
\begin{equation}\label{skj}
s_{kj} = \bigvee_i x_{ij} \land m_{ik}
\end{equation}

\noindent
With the above explaining how we model axon sharing, this formula may be converted to ILP, yielding a formal definition of our solution.
The following indicator variables are used:
$$\forall i,k \in \{1, ..., \text{\# Neurons}\}$$
$$\forall j \in \{1, ..., \text{\# Crossbars}\}$$
$$x_{ij},s_{kj},m_{ik},y_j\in\{1,0\}$$

\noindent
Where:
\begin{itemize}
\item $x_{ij} = 1 \iff$ Neuron $i$ is mapped to crossbar $j$ (output)
\item $m_{ik} = 1 \iff$ Neuron $i$ takes input from neuron $k$
\item $s_{kj} = 1 \iff$ Crossbar $j$ takes neuron $k$ as axonal input
\item $y_j \iff$ Crossbar $j$ is used \textit{at all} in the design
\item $N_j =$ The number of available outputs on crossbar $j$
\item $A_j =$ The number of available inputs on crossbar $j$
\end{itemize}

\vspace*{0.1em} % We need some vpadding here otherwise it looks REALLY weird coming after that list
\noindent % It would be weird to have an indented paragraph here
$N$, $A$, and $m$ are known a priori.
Finally, solutions obey:

\begin{equation}\label{neuron-unique-constraint}
\forall i.\quad \sum_j x_{ij}=1
\end{equation}
\begin{equation}\label{neuron-capacity-constraint}
\forall j.\quad \sum_i x_{ij} \le y_jN_j
\end{equation}
\begin{equation}\label{synapse-constraint-1}
\forall k,j.\quad s_{kj} \le \sum_i x_{ij}m_{ik}
\end{equation}
\begin{equation}\label{synapse-constraint-2}
\forall i.\quad s_{kj} \ge x_{ij}m_{ik}
\end{equation}
\begin{equation}\label{axon-capacity-constraint}
\forall j.\quad \sum_k s_{kj} \le y_jA_j
\end{equation}

Constraint \ref{neuron-unique-constraint} ensures each neuron outputs to one crossbar, while \ref{neuron-capacity-constraint} prevents exceeding crossbar output capacity.
Constraints \ref{synapse-constraint-1} and \ref{synapse-constraint-2} model synapse sharing as expressed in \ref{skj}.
Finally, constraint \ref{axon-capacity-constraint} limits input capacity.
Together, $N_j$ and $A_j$ describe all available crossbar dimensions.
These constraints form the foundation of our mapping algorithm.
\subsection{Optimization for Area}

The formulations in constraints \ref{neuron-unique-constraint}--\ref{axon-capacity-constraint} together describe a valid mapping of an SNN described in $m_{ik}$ for a given MCA described in $A_j$ and $N_j$.
Now, utilizing the solution variables $y_j$, this approach moves from finding \textit{some} valid solution to finding the \textit{best} solution.
The variables $y_j$ describe whether or not crossbar $j$ is "enabled," or has any neurons mapped to it.
Minimizing area is achieved by minimizing the weighted sum of enabled crossbars; a constant area approximation factor $C_j$ is included to consider non-linear area scaling of overhead hardware.
This idea is expressed as the objective:
\begin{equation}\label{area-opt}
min\left(\sum_jy_jC_j\right)
\end{equation}

This objective, together with the earlier constraints, may be passed to an ILP solver to produce solutions optimal in area consumption.
Because of the added complexity of axon sharing and more solution variables ($i \in \{1,...,\#\text{ Neurons}\}$ instead of $i\in\{1,...,\#\text{ MCCs}\}$), optimization will occur more slowly at first compared to SpikeHard.
However, given that SpikeHard can only further improve by being applied iteratively with successively larger MCCs, our approach will overtake it in an acceptable amount of solver time as demonstrated empirically in Section \ref{sec:area_compare}.
\subsection{Static Optimization for Number of Routes}\label{sec:snu}

Modeling axon sharing is not only useful for lower area consumption; accurate counts of axons also allows analyzing the connections between crossbars.
Minimizing the number of these routes has a direct impact on energy consumption, network congestion, and router capability requirements.
The heuristic preferring \textit{local routes} over \textit{global routes} is termed \textit{Static Network Utilization} (SNU), as it provides a static approximation of (chip router) network utilization.

Utilizing the framework provided by earlier constraints, the number of expected total (local+global) network packets is trivial to optimize for:
\begin{equation}\label{synapse-shared}
min\left(\sum_{i,j}s_{ij}\right)
\end{equation}
\noindent
Extending this to count only global routes requires the new variables:
$$b_{kj} \in \{0, 1\}$$
\noindent
Where $b_{kj}$ is 1 if and only if neuron $k$ is used as both output and input on crossbar $j$.
This is realized by the following constraints:
\begin{equation}\label{synapse-opt}
\left.
\begin{aligned}
b_{kj} \ge s_{kj} + x_{kj} - 1\\
b_{kj} \le s_{kj}\\
b_{kj} \le x_{kj}\\
\end{aligned}
\right\}
\quad
\textit{i.e. }b_{ij} = x_{ij} \land s_{ij}
\end{equation}
\noindent
Minimization of global route count is then achieved by counting the number of total routes and subtracting the number of local routes:

\begin{equation}\label{synapse-opt-exec}
min\left(\sum_{i,j} s_{ij}-b_{ij}\right)
\end{equation}
\subsection{Profile-Guided Optimization for Number of Packets}

While SNU minimizes the number of network \textit{routes}, it only approximates the number of network \textit{packets}---the key factor in network congestion and energy use.
By leveraging regularities in SNN structure and application behavior \cite{daniel_j__amit__2003,pastorelli2019scaling,bartram2023parallel}, Profile Guided Optimization (PGO) can be used to penalize frequently used routes more than seldom-used ones.
The result is an improvement in \textit{average} case execution, targeting anticipated network packet count.

Efficiently implementing PGO within an ILP solver for this task requires both simulator support (for dumping profile data) and architectural support.
The following calculations assume the architecture sends only one network packet per crossbar target per neuron fire.
Specifically, networking must respect axon sharing: if neuron $X$ targets both neurons $Y$ and $Z$ within crossbar $j$, only one packet should be generated per spike of $X$.
With this hardware assumption, we introduce the statically known ILP variable $W_i$ representing the profile count of neuron $i$'s spikes during testing.
Thus, anticipated chip router traffic (dynamic network utilization) is minimized by the following objective:

\begin{equation}
min\left(\sum_{i,j} s_{ij}\cdot W_{i}-b_{ij}\cdot W_{i}\right)
\end{equation}

This heuristic will not only perform better by prioritizing frequently used routes but also solve faster. Since many neurons never fire within the profile data, their terms are removed from the above heuristic, enabling the solver to converge towards an optimal solution much more quickly.

\section{Experimental Results}\label{sec:results}

% Typesetting: This figure lives in the area_breakdown section,
% but must be placed here in order to be on the correct page.
% But, that screws with the figure ordering. So we need to hack
% the figure from area_comparison to have the earlier figure
% number. So we artificially increase the figure count by 1
% to skip over the one we're saving for later:
\stepcounter{figure}
% This figure is placed in its own file so that we can move it around
% in the text more easily. Because it's a double-column figure, we must take extra
% care to place it in the right place.

% Tex hacking:
\makeatletter
% \markbb{prefix}{content}
%
% Sets tikzmarks at the bounding box corners of `content'
%
% Tikzmarks are prefixed by `prefix'. For example, Setting prefix to Lastbox
% would create the tikzmarks Lastbox_NW, Lastbox_NE, Lastbox_SW, and Lastbox_SE
%
\newcommand{\markbb}[2]{
\tcbset{markbbstyle/.style={enhanced, boxsep=0pt, left=0pt, right=0pt, top=0pt,
                            bottom=0pt, arc=0pt,
                            boxrule=0pt, toprule=0pt,
                            sharp corners, frame hidden,
                            colback=white,
                            overlay={
                                \begin{scope}
                                    \draw (frame.north west) node {\tikzmark{#1_NW}};
                                    \draw (frame.north east) node {\tikzmark{#1_NE}};
                                    \draw (frame.south west) node {\tikzmark{#1_SW}};
                                    \draw (frame.south east) node {\tikzmark{#1_SE}};
                                \end{scope}
                                }
                            }
}
\begin{tcolorbox}[markbbstyle]
#2
\end{tcolorbox}}
\makeatother

% And it's used as thus:
\begin{figure*}[t]
    \centering
    \begin{minipage}{0.38\linewidth}
        \begin{subfigure}{\linewidth}
            \markbb{BP1}{
                \centering
                % \includesvg[width=\textwidth]{figures/experiments/het_breakdown/out-109-axsh.svg}
                \def\svgwidth{\textwidth}
                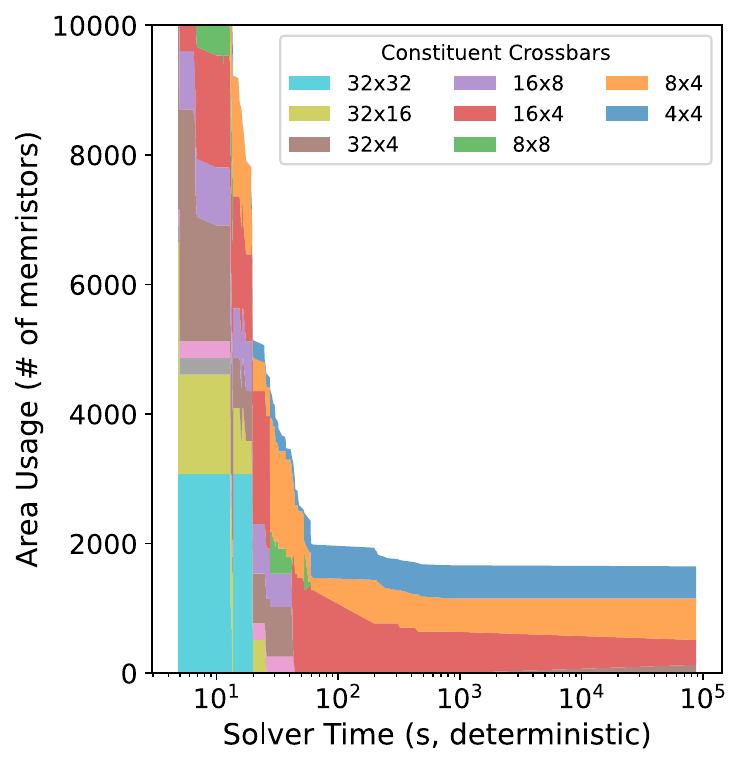
                \caption{Network A Breakdown}
                \label{subfig:breakdown-main}
            }
        \end{subfigure}
    \end{minipage}
    \hfill
    \begin{minipage}{0.6\linewidth}
        \captionsetup[subfigure]{aboveskip=0.049cm}
        \begin{minipage}[t]{\linewidth}
        \centering
        \begin{subfigure}{0.3\linewidth}
            \markbb{BP2}{
                \centering
                \hspace*{-1.3em}
                % \includesvg[width=1.124\textwidth]{figures/experiments/het_breakdown/out-109-axsh-pie.svg}
                \def\svgwidth{\textwidth}
                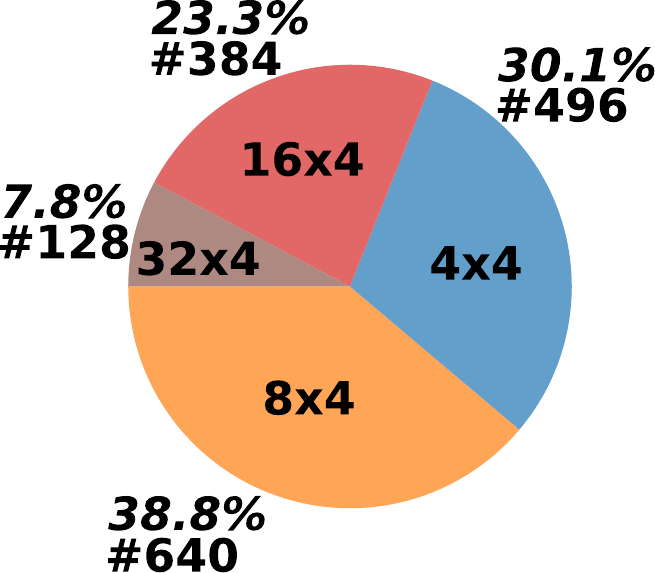
                \caption{Network A}
                \label{subfig:bs-1}
            }
        \end{subfigure}
        \hfill
        \begin{subfigure}{0.3\linewidth}
            \markbb{BP3}{
                \centering
                %\includesvg[width=\textwidth]{figures/experiments/het_breakdown/out-68-axsh-pie.svg}
                \def\svgwidth{\textwidth}
                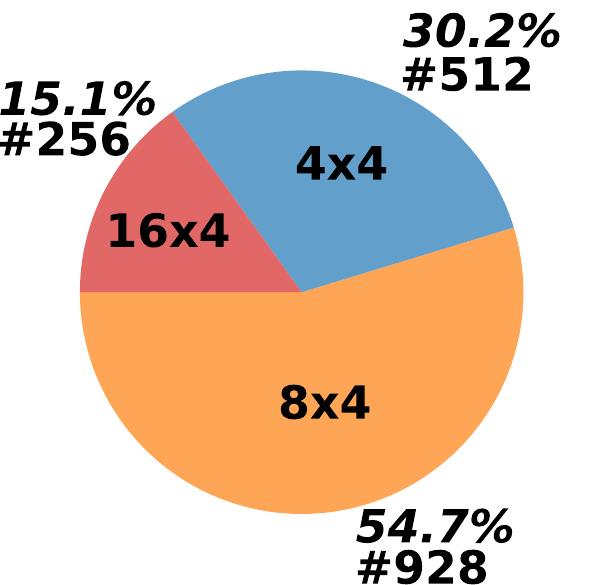
                \caption{Network B}
                \label{subfig:bs-2}
            }
        \end{subfigure}
        \hfill
        \begin{subfigure}{0.3\linewidth}
            \markbb{BP3}{
                \centering
                \hspace*{-0.35em}
                %\includesvg[width=1.04\textwidth]{figures/experiments/het_breakdown/out-42-axsh-pie.svg}
                \def\svgwidth{\textwidth}
                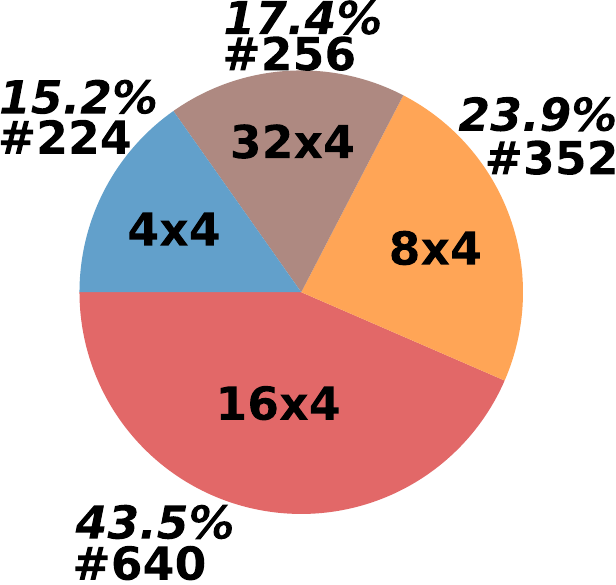
                \caption{Network C}
                \label{subfig:bs-3}
            }
        \end{subfigure}
        \end{minipage}

        \vspace{1.175em} % \vfill doesn't work -_-
        
        \begin{minipage}[b]{\linewidth}
        \begin{subfigure}{0.3\linewidth}
            \markbb{BP4}{
                \centering
                %\includesvg[width=\textwidth]{figures/experiments/het_breakdown/out-144-axsh-pie.svg}
                \def\svgwidth{\textwidth}
                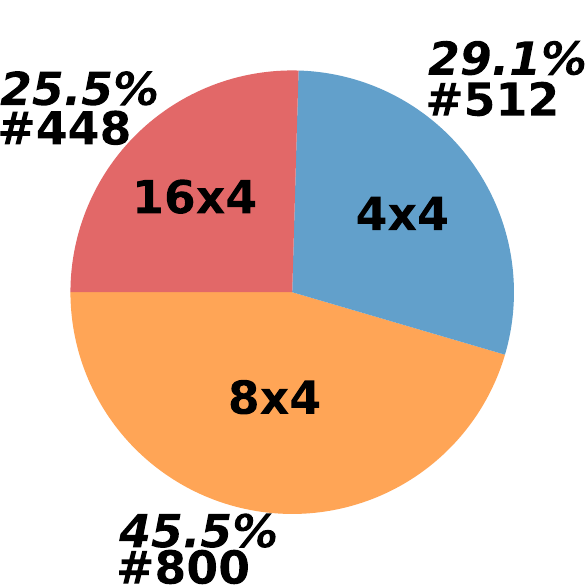
                \caption{Network D}
                \label{subfig:bs-4}
            }
        \end{subfigure}
        \hspace{1.02em}
        \begin{subfigure}{0.3\linewidth}
            \markbb{BP5}{
                \centering
                %\includesvg[width=\textwidth]{figures/experiments/het_breakdown/out-159-axsh-pie.svg}
                \def\svgwidth{\textwidth}
                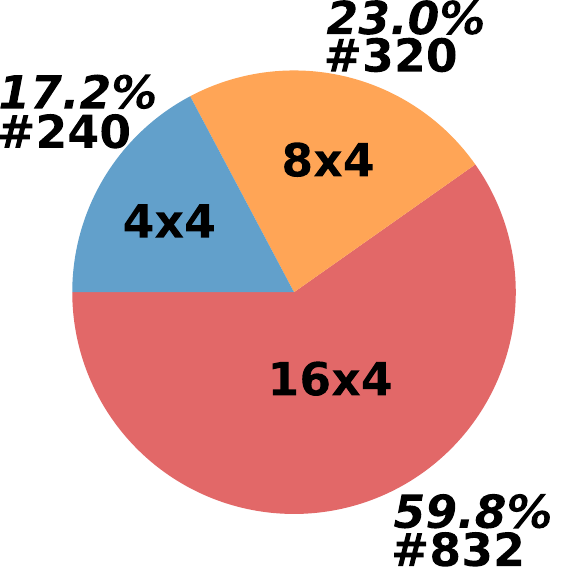
                \caption{Network E}
                \label{subfig:bs-5}
            }
        \end{subfigure}
        \hspace{-1.31em}
        \setlength\tabcolsep{5pt}
        \begin{subfigure}{0.3\linewidth}
            \markbb{BP6}{
                \centering
                \setlength{\tabcolsep}{3pt}
                \renewcommand{\arraystretch}{1.1}
                \begin{tabular}{| l | l |}
                    \hline
                    Network & Best Solution \\
                            & Time (s, det) \\
                    \hline
                    A & 87432\\
                    B & 1637\\
                    C & 122225\\
                    D & 71402\\
                    E & 91257\\
                    \hline
                \end{tabular}
                \caption{Summary}
                \label{fig:sb-summary}
            }
        \end{subfigure}
        \end{minipage}
    \end{minipage}
    
    \begin{tikzpicture}[remember picture, overlay]

        \draw (pic cs:BP2_SE) ++ (0.1, -0.1) coordinate (midsep);
        \draw (pic cs:BP2_NE) ++ (0.25, 0.15) coordinate (figNE);

        % Tikzmarks are weird so we must define complex coordinates like this        
        \path let \p1=(pic cs:BP1_NW),\p2=(figNE) in coordinate (pathstart) at (\x1, {max(\y1,\y2)});
        \path let \p1=(pic cs:BP2_SW),\p2=(midsep) in coordinate (right_join) at (\x1, \y2);
        \path let \p1=(right_join),\p2=(pic cs:BP1_SE) in coordinate (mid_join) at (\x2, \y1);
        \path let \p1=(pathstart),\p2=(midsep) in coordinate (topright) at (\x2, \y1);

        \draw[red, dashed]
        (pathstart)
        -- (figNE)%(topright)
        -- (midsep -| figNE)
        -- (mid_join)
        -- (pic cs:BP1_SE)
        -- (pic cs:BP1_SW)
        -- (pathstart);
    \end{tikzpicture}

    \captionsetup{format=hang}
    \caption{Area optimization targeting reasonable heterogeneous architecture: Dimension (In\thinspace{\small x}\thinspace Out), Area\% and \#Count\\
    The ILP solver did not terminate for these tasks; the best solutions found within a 5 hour limit are reported
    }
    \label{fig:breakdown}

    \vspace*{-1\baselineskip} % IEEETran puts WAY too much padding around figures
\end{figure*}

% And now we decrement to reset back behind the next figure:
\addtocounter{figure}{-2}

\subsection{Selection of Networks}

To emphasize the need for heterogeneous architectures, we selected practical SNNs with high structural sparsity.
These pre-trained SNNs are gathered from recent research \cite{SmartPixelsORNL} identifying properties of particle tracks from high-energy particle collision simulations recorded by next-generation pixel detectors \cite{SmartPixelSensor}.
This topic is of great importance in high-energy physics research and serves as a realistic test case.
Sensor data is converted into spike train format for SNN inference.
Then, a recent version \cite{EonsFinal} of Evolutionary Optimization for Neuromorphic Systems (EONS) \cite{EonsBase} is used to generate and train SNNs within the TENNLab framework \cite{Tennlab}.
Finally, we added simulation support for multi-crossbar neuromorphic processors to the framework.
Table \ref{tab:nets} summarizes attributes of the networks used hereafter.

\begin{table}[b]
\centering
\vspace{-0.54em} % Why is there so much padding
\caption{Attributes of Networks used in Experimentation}\label{tab:nets}
\setlength{\tabcolsep}{4pt}
\renewcommand{\arraystretch}{1.1}
\begin{tabular}{| l | l | l | l | l | l | l |}
\hline
Network & Node  & Edge  & Max    & Edge    & \multicolumn{2}{l|}{Sparsity Index \cite{GiniSparsity}}\\
\cline{6-7}
        & Count & Count & Fan-In & Density & Incoming & Outgoing \\
\hline
% Generated from https://code.ornl.gov/abisko/abisko-software/network-mapper-experiments/-/blob/main/print_net_attributes.py?ref_type=heads
A & 229 & 464 & 11 & 0.0088 & 0.6889 & 0.6764 \\
B & 257 & 464 & 10 & 0.0070 & 0.6411 & 0.6304 \\
C & 148 & 487 & 15 & 0.0222 & 0.5744 & 0.6067 \\
D & 253 & 499 & 13 & 0.0078 & 0.6431 & 0.6541 \\
E & 150 & 446 & 11 & 0.0198 & 0.5876 & 0.6229 \\
% Meaning:
% - 144: D
% - 42: C
% - 159: E
% - 109: A
% - 68: B
\hline
\end{tabular}
\vspace*{0.511em} % Trying to line up with the figure caption
\end{table}

\begin{table}[ht]
\vspace{0.40\baselineskip}
\centering
\caption{Utilized Crossbar Dimensions}\label{tab:size}
\setlength{\tabcolsep}{3pt}
\renewcommand{\arraystretch}{1.1}
\begin{tabular}{| l | l | l | l |}
\hline
Base Dimension & Multi-Macro 2x  & Multi-Macro 4x  & Multi-Macro 8x \\
\hline
4x4   & 8x4   & 16x4 & 32x4 \\
8x8   & 16x8  & 32x8 & --   \\
16x16 & 32x16 & --   & --   \\
32x32 & --    & --   & --   \\
\hline
\end{tabular}
\vspace{-1\baselineskip} % There's an inconsistent amount of space below this table vs the separation between paragraphs 
\end{table}

\subsection{Selection of Crossbar Sizes}
Crossbar size choice is critical to practical optimization, as a natural tradeoff between area and SNU exists at differing crossbar sizes.
For this study, we assume power-of-two square crossbars from 4x4 to 32x32 as supported by \cite{zhou2024recent,huang2024memristor,aguirre2024hardware}.
We reference the multi-macro vertical stacking technique from \cite{NeuroCIM} to supply rectangular crossbars (we assume our square crossbars have the necessary additional capacity to enable this technique).
Crossbars above 32 input channels are excluded, as optimal solutions never included them in preliminary testing.
The total set of allowed crossbar dimensions is shown in Table \ref{tab:size}.

\subsection{Experimental Setup}

We tie the TENNLab framework \cite{Tennlab} to Google's OR-Tools \cite{ORTools}, utilizing the \texttt{SAT\_INTEGER\_PROGRAMMING} solver.
Importantly, Google OR-Tools exposes \textit{deterministic} timing results reflecting only the number, type, and complexity of each solver operation.
Deterministic timing closely approximates wall clock time if unlimited resources (cores, memory, etc) are present.
To evaluate our method independent of scaling potential, we provide only deterministic timing.

Crossbar sizes are either 16x16 (the smallest power-of-two size capable of fitting the most fan-in intense networks from Table \ref{tab:nets}) for homogeneous experiments or pulled from Table \ref{tab:size} for heterogeneous experiments.

\subsection{Area Comparison}\label{sec:area_compare}

Fig. \ref{fig:area} reports area consumption reduction per utilized solver runtime.
Although adding per-unit area overhead is supported, we only consider memristor count to focus on the effectiveness of our method absent of hardware specifics.
Four configurations are tested: with MCC or axon-sharing, targeting homogeneous or heterogeneous configurations.
Improvement is relative to each network's best result under MCC targeting a homogeneous MCA.
SpikeHard was applied repeatedly until convergence was achieved.

\begin{figure}[H] % This WILL fit on the same page
    \captionsetup{skip=2pt}
    \vspace{-0.2\baselineskip} % Even if it's the last thing I do
    \centering
    %\includesvg[width=1\linewidth]{figures/experiments/area.svg}
    \def\svgwidth{1\linewidth}
    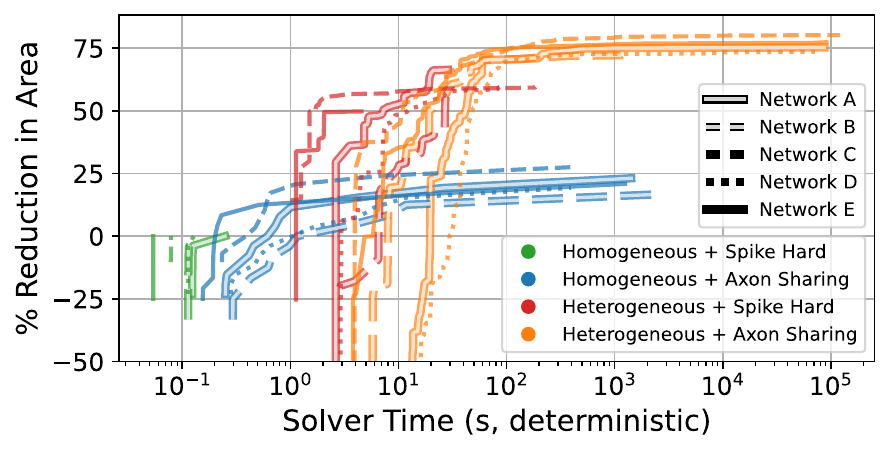
    \caption{Relative Improvements in Area Optimization}
    \label{fig:area}
    \vspace{-0.001\baselineskip} % AAAAAAAAAAAAA
\end{figure}

The results indicate that modeling axon sharing reduces area \rangeof{area_hom} more than SpikeHard for homogeneous MCAs, though \rangeof{breakeven_hom} additional solver time is needed to break even.
A clear preference for heterogeneous MCAs is seen, with axon sharing providing \rangeof{area_het_over_hom} \textit{further} area reduction.
Overhead is also lower in the heterogeneous case, requiring just \rangeof{breakeven_het} more solver time to break even.

\begin{comment}
Summary:\\
- SH Hom -> Hom:  16.666666666666664-27.586206896551722\% \\
- SH Hom -> SH Het:  49.73958333333333-66.34615384615384\% \\
- SH Hom -> Het:  72.39583333333334-80.17241379310344\% \\
- Hom -> Het:  66.875-72.61904761904762\%
\end{comment}

% And finally, now that we have the area comparison figure
% drawn, we can fix the counter:
\addtocounter{figure}{2}
% And there we go. Now the figures are numbered in the order
% they are logically referenced in the document.
% We're now free to include the text for area_breakdown
% (which does not have the figure):
\subsection{Area Breakdown}

By plotting every intermediate solution, we explore \textit{how} OR-Tools refines solutions.
Fig. \ref{fig:breakdown} shows such results for all networks in the study, focusing on one particular network in subfigure \ref{subfig:breakdown-main}.
Although the plot is clipped for early values, preferred crossbar sizes were clearly identified quickly before solutions were slowly refined.
Despite subfigure \ref{fig:sb-summary} showing high \textit{best} solution times, all networks exhibited the same trend of finding \textit{near-best} solutions quickly.
Subfigures \ref{subfig:bs-1}--\ref{subfig:bs-5} summarize the best solutions found.
Despite the availability of larger crossbar sizes, a clear trend towards taller crossbars emerged due to the structural sparsity of the input SNNs.

While not explored further in this paper, these lessons could guide further research toward finding optimal solutions more quickly.
For example, the iterative swapping approach in \cite{RecentApproach1} is validated with our data.

% The actual include is somewhere else in the document.
% This is required to get this figure on the correct page.
% \input{figures/experiments/breakdown_figure}

% And the rest of the section:
\subsection{Static Network Utilization}

To explore how the SNU optimization from Section \ref{sec:snu} minimizes the number of routes, we took the area-optimal solutions from the previous experiment, restricted the set of enabled crossbars to not increase area, and optimized for SNU.
Fig. \ref{fig:snu-improvement-hom} shows results for the homogeneous case, and Fig. \ref{fig:snu-improvement} for the heterogeneous case.
Both cases show similar improvements: \rangeof{snu_hom} for homogeneous and \rangeof{snu} for heterogeneous.
Improvement is reported relative to the most area-optimal solution found by the solver.

\begin{figure}[H]
    \vspace{-0.23em} % Phantom out internal svg padding
    \centering
    %\includesvg[width=1\linewidth]{figures/experiments/snu_hom.svg}
    \def\svgwidth{1\linewidth}
    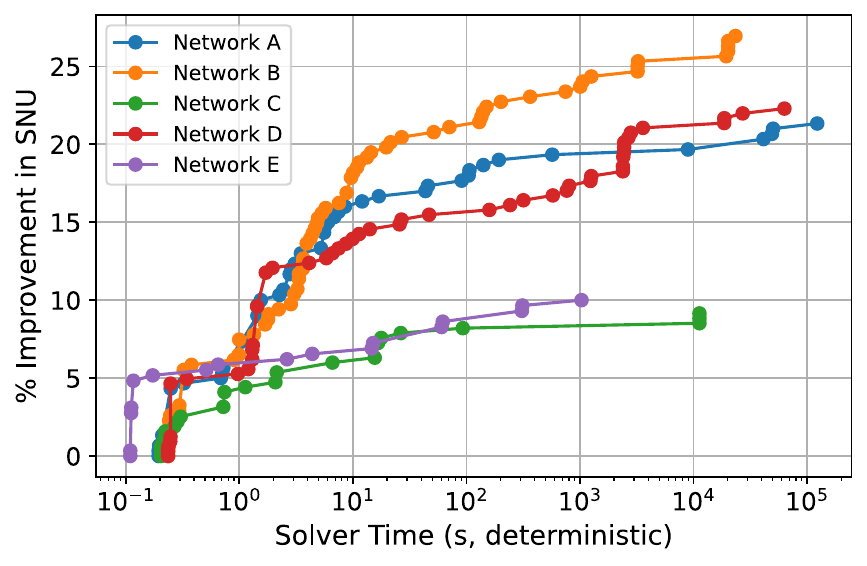
    \captionsetup{format=hang}
    \caption{Optimization of Routes over Already Area-Optimal Solutions for Homogeneous Architecture}
    \label{fig:snu-improvement-hom}
\end{figure}
\vfill
\begin{figure}[H]
    \vspace{-3em} % Why is there so much padding
    \centering
    %\includesvg[width=1\linewidth]{figures/experiments/snu.svg}
    \def\svgwidth{1\linewidth}
    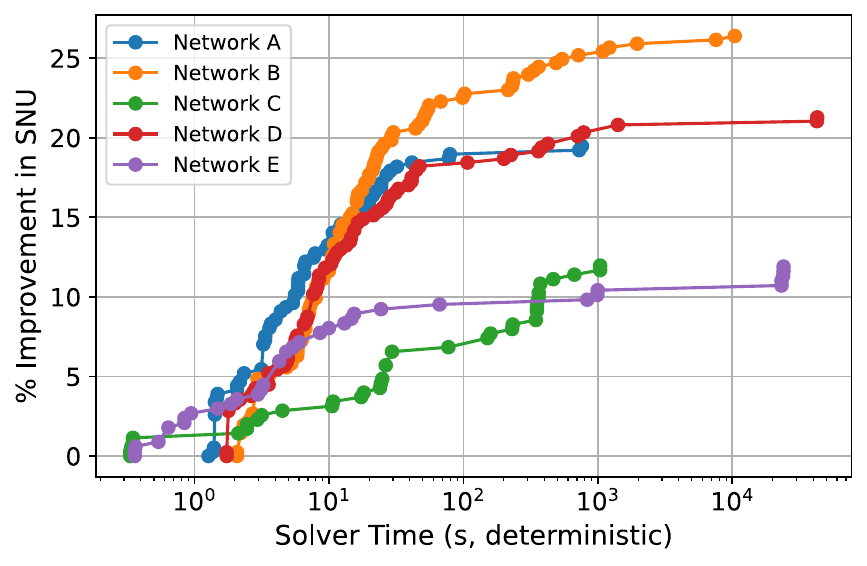
    \captionsetup{format=hang}
    \caption{Optimization of Routes over Already Area-Optimal Solutions for Heterogeneous Architecture}
    \label{fig:snu-improvement}
    \vspace{-0.415em} % Go the the bottom of the page please
\end{figure}

\subsection{Area-SNU Evolution}

% The first paragraph of this section is to be typeset
% sandwitched between the figures

\begin{figure}[t]
    \centering
    \includegraphics[width=1\linewidth]{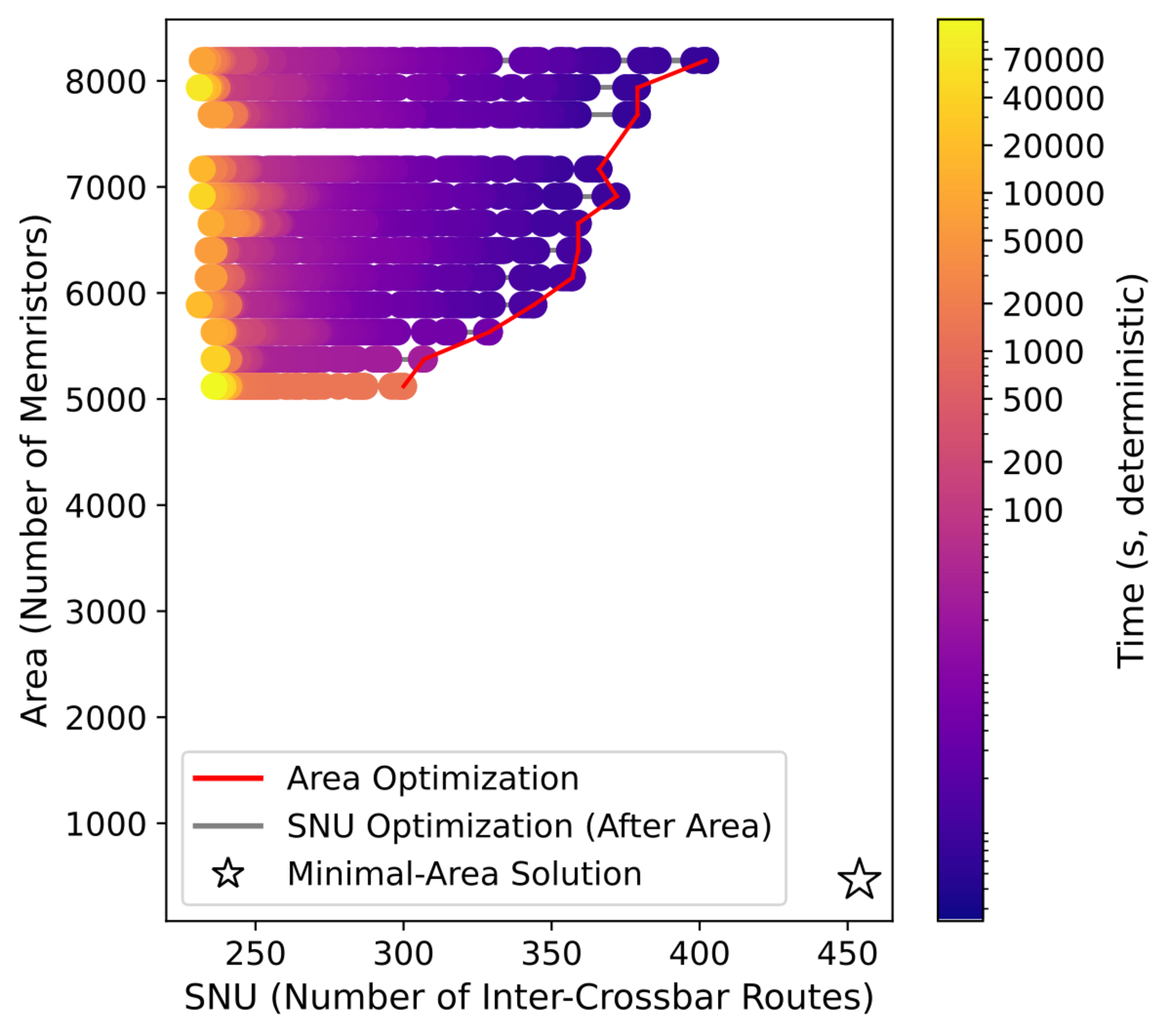}
    \captionsetup{format=hang}
    \caption{Area/SNU Evolution for Network A Targeting\linebreak Homogeneous MCA}
    \label{fig:evo-hom}
    \vspace{-\baselineskip} % There's an unreasonable amount of padding
\end{figure}

\begin{figure}[b]
    \vspace{-0.75\baselineskip} % Here too
    \centering
    \includegraphics[width=1\linewidth]{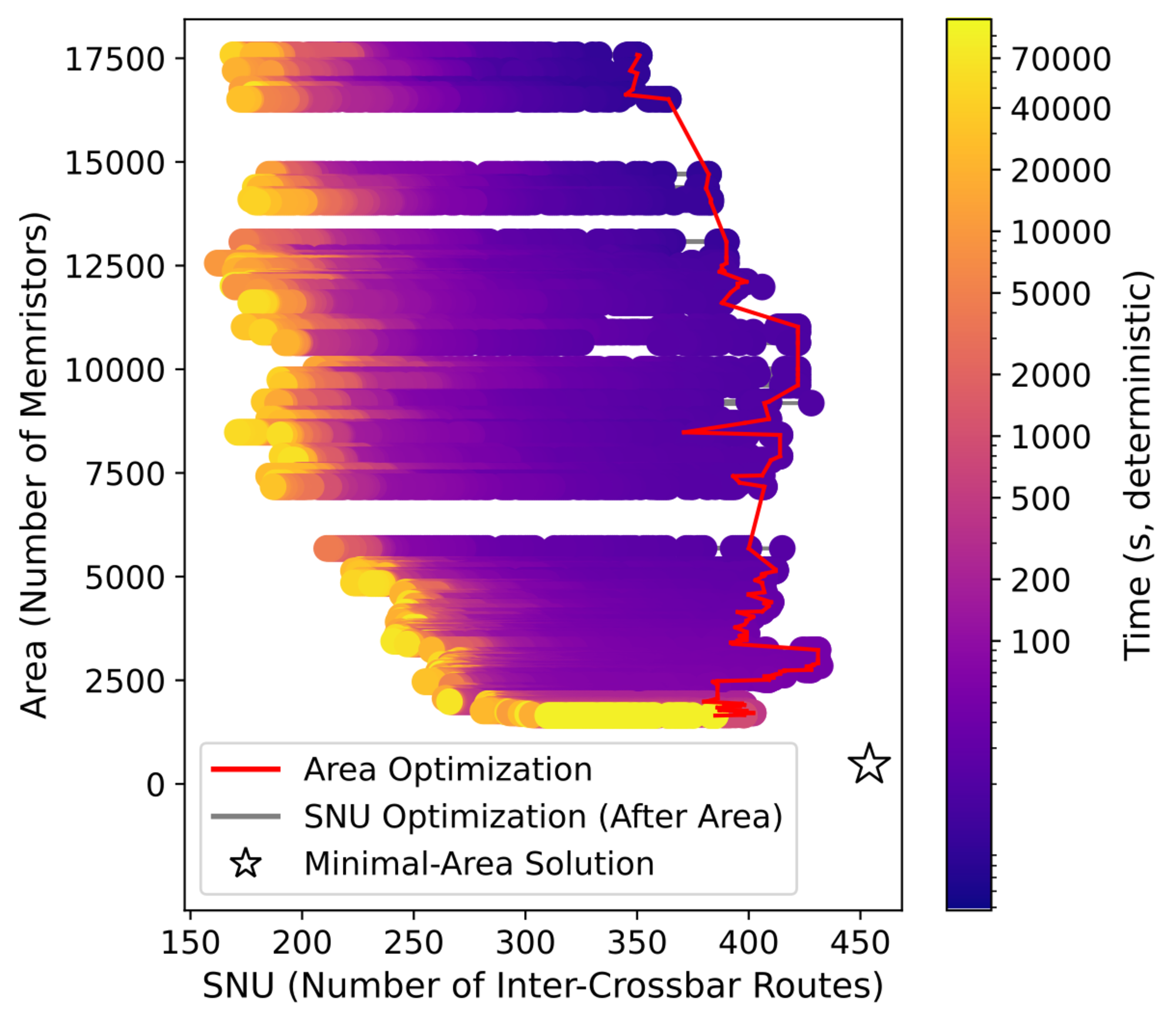}
    \captionsetup{format=hang}
    \caption{Area/SNU Evolution for Network A Targeting\linebreak Heterogeneous MCA}
    \vspace{-0.4em} % And for some ??? reason this figure isn't [b]
    \label{fig:evo-het}
\end{figure}

\looseness-1
While the area and SNU optimization results from the previous section are promising, they do not fully capture the trade-off between the two optimizations.
To illustrate this, we selected one network for further analysis.
In Fig. \ref{fig:evo-hom}, area optimization for the homogeneous architecture was performed, with every intermediate solution forming the basis for SNU optimization.
The \textit{total} solver time is reported, along with a mark indicating where a hypothetical minimal-area solution of one neuron per minimally sized crossbar would fall.
While not achievable in any target architecture of this study, this point communicates a bound on the solution space.

Similarly, Fig. \ref{fig:evo-het} shows the results for the heterogeneous case.
Although early solutions are less optimal due to added hardware complexity, uniform improvements over the homogeneous case in area, power, and solver time are made quickly.
%At the optimization limit, a trade-off emerges between the two metrics---a trend consistent across all networks in this study.
At the optimization limit, a trade-off emerges between the two metrics; this trend is sensitive to the target architecture and consistent across all networks in this study.
\subsection{Dynamic Network Utilization}

\begin{figure}[t]
    \setlength{\abovecaptionskip}{3pt}
    \vspace{-0.03\baselineskip} % Trying to align through the colorbar on Fig. 7
    \centering
    % \includesvg[width=1\linewidth]{figures/experiments/pgo.svg}
    \def\svgwidth{1\linewidth}
    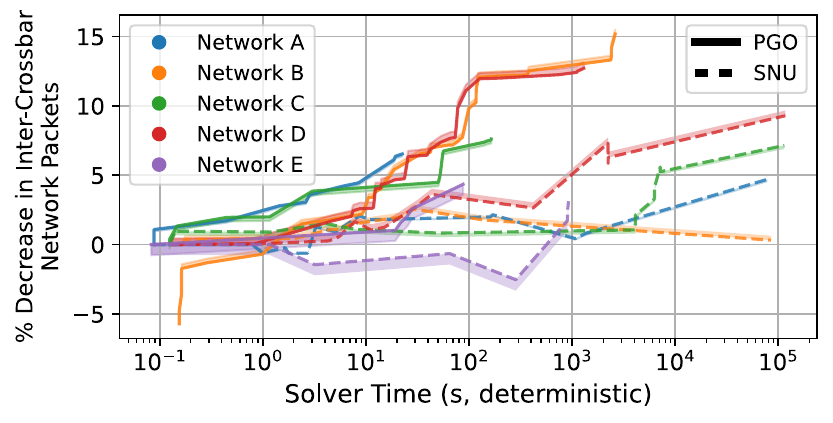
    \caption{Profile-Guided vs Static Optimization}
    \label{fig:pgo}
    \vspace{-1.1\baselineskip} % There's an unreasonable amount of padding
\end{figure}

\looseness-1
This final experiment briefly showcases the profile-guided version of SNU.
Instead of minimizing the number of routes, the number of network packets are minimized according to spike profile data.
This data includes SmartPixel simulations of high-energy particle collisions \cite{SmartPixelSensor}---the same data used to train and evaluate the networks in this study.
A randomly-selected 1\% sample of the data (51MB) was used for PGO, with optimization results shown in Fig. \ref{fig:pgo}.
This figure includes error bands indicating spike count under execution of the other 99\% of the data (5.0GB) within the same application.

The reported results indicate \rangeof{pgo} decrease in spike count compared to the best SNU-optimized networks while requiring \rangeof{pgo_magnitude_speedup} orders of magnitude less solver time.
Additionally, due to the low error, the results confirm that spiking activity is regular enough to benefit from PGO.

\section{Conclusion}
This paper addresses growing sparsity in Spiking Neural Networks (SNNs) and the resulting opportunity to reduce area consumption through heterogeneous Memristor Crossbar Architectures (MCAs).
By developing Integer Linear Programming (ILP) formulations supporting heterogeneously sized crossbars and optimizing axonal interactions, we significantly outperform previous methods in area efficiency.
We show a \rangeof{area_hom} reduction in area for homogeneous MCAs and a substantial \rangeof{area_het_over_hom} \textit{further} reduction for heterogeneous MCAs.
Additionally, we introduce an optimization to minimize inter-crossbar routing, achieving an \rangeof{snu} reduction without increasing area.
Finally, we propose a profile-guided approach to reduce inter-crossbar spike count \rangeof{pgo} more than the best route-minimized solutions, while requiring \rangeof{pgo_magnitude_speedup} orders of magnitude less solver time.
These contributions demonstrate the potential of heterogeneous MCAs in SNN acceleration, enabling progress toward more efficient and scalable neuromorphic hardware.

\clearpage

\bibliographystyle{IEEEtran}
\bibliography{references,vetter}

\phantom{a}

}

\end{document}